# Autocatalytic knowledge dynamics in the AI era

Boris M. Dolgonosov
Independent Researcher, Haifa, Israel; borismd31@gmail.com

**Abstract.** In the AI era, knowledge is growing at an accelerated rate. The goal of this study is to construct the macrodynamics of knowledge and elucidate the role of AI in this process. An autonomous knowledge production system is examined. The driving force behind the system's development is the disturbance of its homeostasis under the impact of external or internal factors. To solve emerging problems, the system uses available or generates new knowledge. The process is autocatalytic in the sense that knowledge accumulated over the past is used to create new knowledge. Knowledge production is hampered by resource scarcity and environmental disturbances that worsen the system's state. The loss of knowledge, in particular due to its obsolescence, contributes to the process' slowdown. Information noise also inhibits the advancement of knowledge, leading to energy dissipation. The combination of autocatalysis, aimed at generating new knowledge, with counteracting forces caused by resource scarcity, environmental problems, knowledge loss, and information noise provides the basis for building knowledge dynamics. The resulting dynamic equation shows that, depending on the balance of the forces acting, the rate of knowledge production can change as follows: 1) increase, ensuring system's development; 2) decrease to a finite (non-zero) level: the system degrades but survives; and 3) decline to zero: the system collapses. The model was applied to WIPO patent filing data, yielding parameter values. The influence of AI enhances autocatalysis, but counteracting factors also increase. A shift in the balance of forces leads to a phase transition with a change in knowledge dynamics. The system's evolution can proceed through a series of phase transitions. Key milestones in AI development are compared with the forces represented in knowledge dynamics. Barriers to knowledge growth are discussed.

## 1. Introduction

A civilization can be viewed as an intelligent system endowed with the following attributes:

- *a world picture* consisting of a multitude of functionally interconnected verified information objects;
- *a hierarchy of goals*, the achievement of which is necessary to maintain the system's homeostasis;
- *methods of information processing*: an object obtained as a result of targeted processing of incoming information, after verification and binding with other objects, is integrated into the world picture and becomes a component of knowledge.

The epistemological characteristic “justified true belief” (Mugleston et al., 2025) applies to objects included in the world picture.

The level of development of an intelligent system is assessed by the volume of knowledge. From the perspective of knowledge macrodynamics, this system can be viewed as a continuum with certain macrocharacteristics, without detailing its internal structure.

The world picture can be organized as a knowledge base with connections between objects. Then the measure of knowledge will be the total memory space occupied by objects in the knowledge base.

The need to increase knowledge can be driven by various factors. It can be a purely scientific need to develop a theoretical problem or a practical need to solve a pressing problem. This creates a driving force that compels investments in solving these problems. The scale of investment is an indicator of the problem's urgency and complexity.

A problem-solving method is developed based on existing knowledge. Once verified and integrated into the world picture, the resulting method becomes new knowledge. Thus, a positive feedback loop emerges: *knowledge creates knowledge*. This *autocatalysis* significantly influences the rate of knowledge production, which, in the absence of obstacles, grows according to a hyperbolic law. The peculiarity of this law is that in the early stages, when knowledge is extremely small, development is virtually imperceptible. But in the later stages, when sufficient knowledge has accumulated, development accelerates so much that a singularity point is reached within a short time.

In reality, singularity cannot be achieved due to resource limitations, primarily energy. Knowledge production also requires infrastructure that has been built over a long time and that has been supported using energy generated throughout history. The current functioning of the infrastructure is accomplished by energy produced at present. Global GDP is determined entirely by two factors: current energy intensity and energy produced in the past and materialized in infrastructure (Dolgonosov, 2020). Expressing GDP in terms of energy establishes a monetary equivalent for energy. Investments in knowledge cannot exceed a certain percentage of GDP. This is one of the constraints on the rate of knowledge production imposed by resource limitations.

The environment also imposes limitations. Energy production generates heat, which increases with civilization's energy output. At high output, atmospheric overheating is inevitable even with a transition to carbon-free energy, which only temporarily alleviates the problem. A radical solution is to move energy-intensive production off-planet. This creates an environmental limitation on energy production.

Thus, the rate of knowledge production is directly or indirectly limited by a certain level of energy generation. The impossibility of increasing energy consumption indefinitely slows down knowledge production and smooths out the hyperbolic singularity.

Knowledge production is complicated by a reverse process—knowledge loss, which occurs when obsolete ideas, theories, and technologies fall out of circulation. Not everything physically disappears; a significant portion of knowledge remains in archives, but becomes unused and passive. If necessary, passive knowledge can be restored with appropriate funding, but then fewer resources remain available for investment in the production of new knowledge.

Knowledge production is further complicated by the mass generation of unreliable or even deliberately false information online. This creates noise that drowns out useful information. The need to filter out information noise also diverts investment from knowledge production. New results become bogged down in this noise: energy is expended promoting them in a chaotic environment, delaying their verification and incorporation into the world picture. Information noise leads to energy dissipation. This process is analogous to a well-known physical phenomenon: resistance to the motion of a particle in a viscous medium. This analogy proves useful when considering the dynamics of knowledge.

The development of AI accelerates the production of new knowledge. In accordance with the set goal, AI provides rapid access to accumulated knowledge, uses appropriate information processing methods, tests the obtained information objects, integrates them into the world picture by establishing connections with other objects, and thereby converts the obtained objects into new knowledge. Human participation in this process is implied, although his degree of involvement may decrease as AI develops, shifting to the area of goal formulation and problem setting. The nature of human interaction with AI is conveyed by the following quote: “Through recursive feedback, where humans shape AI, and AI increasingly shapes human thought and action, AI may acquire a role not as a separate agent, but as a core architectural element of an emerging collective individual” (Rainey and Hochberg, 2025).

An overview of approaches directly or indirectly related to knowledge production is given in the work (Dolgonosov, 2024).

The above provides the basis for constructing the macrodynamics of knowledge, which is the subject of this work. The sequence of presentation is briefly as follows:

- the equation for knowledge dynamics is substantiated;
- based on this, various modes of knowledge production are analyzed;
- the application of the developed model to certain specific cases is considered; and
- the role of AI in knowledge production within the model is discussed.

## 2. Evolution of an intelligent system

### *2.1. Master equation*

Let us consider the evolution of a knowledge-production system. Let $q(t)$ be the volume of knowledge accumulated in the system by time $t$. The rate of knowledge production is defined as

$$v = \frac{dq}{dt}. \tag{1}$$

Knowledge production is *an autocatalytic process* in which existing knowledge generates new knowledge. Therefore, the dynamic equation describing the acceleration of knowledge production $dv/dt$ must contain a quadratic term $av^2$, where $a$ specifies the intensity of autocatalysis. Autocatalysis, described by the equation $dv/dt = av^2$, leads to *a hyperbolic growth* of the rate of cognition $v = v_0(1 - t/t_a)^{-1}$ (where time is counted from zero and rate from $v_0$), in which singularity is reached in a finite time—at the moment $t_a = 1/(av_0)$. Hyperbolic growth reflects the "network effect" of knowledge. Each new discovery is not simply added to the pool, but establishes connections with everything that has come before, generating a cascade of new hypotheses. This is precisely what we are seeing today in AI, which is expanding our ability to *handle complexity*.

As singularity approaches, resource limitations begin to be felt, leading to a transition from unbridled growth to *a logistical constraint* [similar to the Verhulst equation (Cramer, 2002)], which inhibits knowledge production. These inhibitory factors can be divided into two groups.

The first group includes resource scarcity and environmental limitations. These factors can be combined and represented as *an information equivalent b*—a resource-environmental brake whose reciprocal value, $1/b$, limits the global rate of knowledge production. This brake is formalized as a factor $(1 - bv)$, by which the autocatalytic term is multiplied, yielding $dv/dt = av^2(1 - bv)$. Over time, such a system, after hyperbolic growth, reaches a level $v = 1/b$, which represents *a planetary barrier*.

The resource-environmental brake generally describes the physical limits of the world system—the lack of material and human resources, finance, energy, computing power, as well as a wide range of environmental constraints. This problem is especially pressing with the development of AI, which utilizes enormous amounts of resources, providing *structured knowledge* in exchange.

A related constraint is that large-scale computing converts electricity into heat in significant quantities. As AI-powered knowledge production reaches a planetary scale, the limitation becomes not only energy consumption but also the planet's ability to dissipate the generated heat to prevent local or global atmospheric overheating.

This is one of the manifestations of environmental limitations. Their role is to maintain *the planet's technological metabolism* at a safe level. As knowledge develops, *intelligence inevitably strives toward a state where it becomes the driving force commensurate with the power of nature*.

The second group of inhibitory factors considers the complexity of processing and implementing new knowledge, as well as the inactivation and loss of knowledge for the following reasons:

- diversity of approaches to the search for new knowledge, some of which are far removed from reality and act as informational noise, forcing us to waste resources and time filtering it. *Trust* and *verification* are becoming scarce resources. AI can generate a large number of hypotheses in a short time, but humanity will have to spend many years just to verify which ones won't lead to disaster;
- difficulties in implementing verified new knowledge into practice, for example, due to psychological reasons. Knowledge has difficulty making its way into a society that resists innovation— like a viscous medium that impedes mechanical movement;

- inactivation of knowledge due to its obsolescence. Knowledge does not necessarily disappear completely: it can be stored on physical media, but remain unclaimed and passive. Returning it to an active state requires funding and time;
- intentional destruction of information and dissemination of false information as a result of competitive relations and social conflicts, which creates chaos in the information sphere.

All these factors slow down progress towards new knowledge and add a negative term $(-cv)$ to the dynamic equation, where $c$ is the coefficient of knowledge loss due to the combined effect of the listed factors.

The above leads to the construction of *the master equation of autocatalytic knowledge dynamics*:

$$\frac{dv}{dt} = a(q)v^2[1 - b(q)v] - c(q)v, \tag{2}$$

where $a, b,$ and $c$ are coefficients of autocatalysis, resource-environmental braking, and knowledge loss, respectively. The main problem of AI is to increase the rate of knowledge production by increasing the autocatalysis coefficient $a(q)$, as well as reducing the inhibitory factors $b(q)$ and $c(q)$.

Let us introduce the quantity

$$v_b = \frac{1}{b}, \tag{3}$$

which we will call *the threshold rate* (as noted above, this is the planetary barrier). Now the expression in square brackets in (2) can be written as $(v_b - v)/v_b$. This shows that the transition of the rate $v$ through the threshold $v_b$ changes the sign of this expression, and therefore the sign of the entire first term on the right side of equation (2). For example, if $v$ becomes greater than $v_b$, then instead of knowledge production, we will experience a loss of knowledge due to a critical shortage of resources or dangerous environmental disturbance. As the resource-environmental brake $b$ increases, the threshold rate $v_b$ decreases, which further limits the rate of knowledge production $v$.

Let the system have knowledge $q_0$ at the initial time $t_0$. For evolution to begin, the rate of knowledge production must be non-zero: $v_0 > 0$. The corresponding initial condition is

$$t = t_0, \;\; q = q_0, \;\; v = v_0. \tag{4}$$

The coefficients $a$, $b$, and $c$ in equation (2) generally depend on the system's history. This refers to the situation in which the development of an intelligent system is considered not from its inception (when knowledge does not yet exist: $q = 0$), but from some finite moment $t_0$, when a certain volume of knowledge already exists: $q = q_0 > 0$. In this situation, the coefficients $a$, $b$, and $c$ will depend not on the difference $q - q_0$ (i.e., what has been produced since $t_0$), but on all knowledge $q$ accumulated over the entire history of the system. An exception is a useful idealization in which the coefficients are assumed to be constant, and then they do not depend on $q$ at all.

*2.2. Autonomous systems*

Let us consider an autonomous system that produces knowledge. Autonomy implies the absence of an external source of knowledge (from another civilization). AI is not an external source, but an internal entity, regardless of whether it acts as a tool or a creator. AI already, in some cases, realizes itself as a hypothesis generator and solves problems that are insurmountable for humans [for example, decoding 200 million proteins (AlphaFold, 2026) or discovering new antibiotics (Barron, 2025)]. Essentially, *we are now experiencing a phase transition in which AI ceases to be just a calculator—it becomes the planet's intelligent infrastructure that is autonomous*, and in an autonomous system, the rate $v$ and coefficients $a$, $b$, and $c$ in equation (2) do not depend on time explicitly, but only through $q(t)$.

In the absence of an explicit dependence on time, the following relationship holds:

$$\frac{dv}{dt} = v\frac{dv}{dq}, \tag{5}$$

which allows us to reduce equation (2) for the function $v(q)$ to the form

$$\frac{dv}{dq} = a(q)v[1 - b(q)v] - c(q). \tag{6}$$

From here we can find $v(q)$, and then from (1) – the accumulation of knowledge $q(t)$ by time $t$.

## 3. Decomposition of the system

The intelligent system can be divided into two subsystems: fast and slow. *The fast subsystem* manages to reach the threshold rate $v_b$ before the coefficients $a$, $b$, and $c$ change noticeably, i.e., they can be considered constant.

When considering long-term dynamics, slow processes must be taken into account. It is *the slow subsystem* that is responsible for the gradual adaptation of model coefficients to the accumulation of knowledge.

This decomposition of the system allows the use of *an adiabatic approximation*, according to which, at the first stage, the problem is solved for the fast subsystem with constant coefficients, and at the second stage, their slow change is considered, which is introduced into the solution obtained at the first stage.

On the contrary, if the coefficients $a$, $b$, and $c$ change as rapidly as the rate of knowledge production $v$, the above decomposition of the system becomes impossible. Rapid development of AI near the singularity point can lead to a situation where autocatalysis occurs with high intensity. In this case, the dynamic equation must be solved with the coefficients depending on knowledge volume $q$. The nature of this dependence is discussed in Section 5.1. Numerically solving this problem is straightforward. However, the problem lies in the large number of parameters in (24)–(26), which can vary over wide ranges. Calibrating this model requires representative data series, which do not exist yet.

Next, we will consider a situation where decomposition is possible, and analyze the fast and slow subsystems separately.

## 4. Fast subsystem

### *4.1. Dynamics with constant model coefficients*

Consider the case of constant coefficients $a, b, c > 0$, and write (6) in the form

$$\frac{dv}{dq} = -a\left(bv^2 - v + \frac{c}{a}\right),\ \ q > q_0,\ \ v(q_0) = v_0. \tag{7}$$

Let us normalize equation (7) by introducing dimensionless variables $\tau, x, y$ and parameters $\mu, y_0$:

$$\tau = \frac{t - t_0}{t_s},\ \ x(\tau) = a[q(t) - q_0],\ \ y(x) = bv(q), \tag{8}$$

$$t_s = \frac{b}{a},\ \ \mu = \frac{bc}{a},\ \ y_0 = bv_0, \tag{9}$$

where $t_s$ is the characteristic time scale, $\mu$ is the ratio of the inhibitory factors (resource-environmental brake and knowledge loss) to the intensity of autocatalysis, $y_0$ is the ratio of the initial rate of knowledge production $v_0$ to the threshold rate $v_b = 1/b$.

The initial value problem (7) in dimensionless form is

$$\frac{dy}{dx} = -(y^2 - y + \mu),\ \ x > 0,\ \ y(0) = y_0 > 0. \tag{10}$$

Normalization reduces the number of parameters from six $(a, b, c, t_0, q_0, v_0)$ to two $(\mu, y_0)$.

*4.2. Knowledge production rate*

The discriminant of the right side of equation (10) is

$$D = 1 - 4\mu. \tag{11}$$

Depending on the value of $\mu$, which determines the discriminant sign, three different dynamic modes are possible. Integration of equation (10) leads to the following results for each mode:

**Mode 1**. Weak inhibition and strong autocatalysis ($4bc < a$):

$$0 \le \mu < \frac{1}{4},\ \ D > 0: \qquad y = \frac{y_+ - y_- A e^{-kx}}{1 - A e^{-kx}}, \tag{12}$$

where

$$A = \frac{y_0 - y_+}{y_0 - y_-},\ \ y_\pm = \frac{1 \pm k}{2},\ \ k = \sqrt{D}. \tag{13}$$

There are three ranges for $y_0$ with different dynamics, corresponding to low ($y_0 < y_-$), medium ($y_- < y_0 < y_+$), and high ($y_0 > y_+$) initial rate of knowledge production. If $\mu = 0$, then $y_- = 0$, and the first range for $y_0$ disappears. Values at the boundaries of the ranges ($y_0 = y_-$ and $y_0 = y_+$) yield the trivial solution $y(x) = y_0$, which is of no interest.

**Mode 2**. Inhibition and autocatalysis are balanced ($4bc = a$):

$$\mu = \frac{1}{4},\ \ D = 0: \qquad y = \frac{1}{2} + \frac{y_0 - \frac{1}{2}}{1 + \left(y_0 - \frac{1}{2}\right) x}. \tag{14}$$

The dynamics are different for $y_0 < 1/2$ and $y_0 > 1/2$. For $y_0 = 1/2$, the solution is trivial: $y(x) = y_0$.

**Mode 3**. Strong inhibition and weak autocatalysis ($4bc > a$):

$$\mu > \frac{1}{4},\ \ D < 0: \qquad y = \frac{1}{2} - C\,\mathrm{tg}(Cx - \theta), \tag{15}$$

where

$$\mathrm{tg}\,\theta = \frac{2y_0 - 1}{\sqrt{-D}},\ \ C = \frac{\sqrt{-D}}{2}. \tag{16}$$

*4.3. Knowledge accumulation*

The accumulation of knowledge $q(t)$ occurs according to equation (1) with the initial condition (4). In dimensionless form, this looks like this:

$$\frac{dx}{d\tau} = y(x), \quad x(0) = 0. \tag{17}$$

Substituting the expressions for $y(x)$ from (12), (14) or (15) into equation (17) and integrating, we obtain the following results for each mode:

**Mode 1.** $D > 0$:

$$\frac{x}{y_+} - \frac{1}{2\mu} \ln \frac{1 - pe^{-kx}}{1 - p} = \tau, \tag{18}$$

where

$$p = \frac{y_0 y_- - \mu}{y_0 y_+ - \mu}. \tag{19}$$

**Mode 2.** $D = 0$:

$$2x - 4 \ln\left(1 + \left(1 - \frac{1}{2y_0}\right)\frac{x}{2}\right) = \tau. \tag{20}$$

**Mode 3.** $D < 0$:

$$x - 2 \ln \frac{\sin(Cx + \varphi)}{\sin \varphi} = 2\mu\tau, \tag{21}$$

where

$$\operatorname{tg} \varphi = \frac{2C}{y_0 - 2\mu}. \tag{22}$$

The transition from one mode to another is a phase transition, which changes the system's dynamics. This is illustrated by the phase diagram in Fig. 1.

The area *inside the parabola* ($0 \leq \mu < 1/4$) corresponds to an increase in the rate of knowledge production, reaching a plateau. On the phase plane *above the parabola*, the rate decreases to a finite positive level. In both cases, the system survives. *Below the parabola*, the rate decreases to zero, which is fatal for the system.

At $\mu = 1/4$, corresponding to the parabola's apex, the upper point represents a scenario in which the rate decreases to a finite limit, so the system survives. At the lower point, the rate reaches zero, indicating the system's collapse.

Farther from the parabola ($\mu > 1/4$), the scenarios are always fatal: the rate drops to zero.

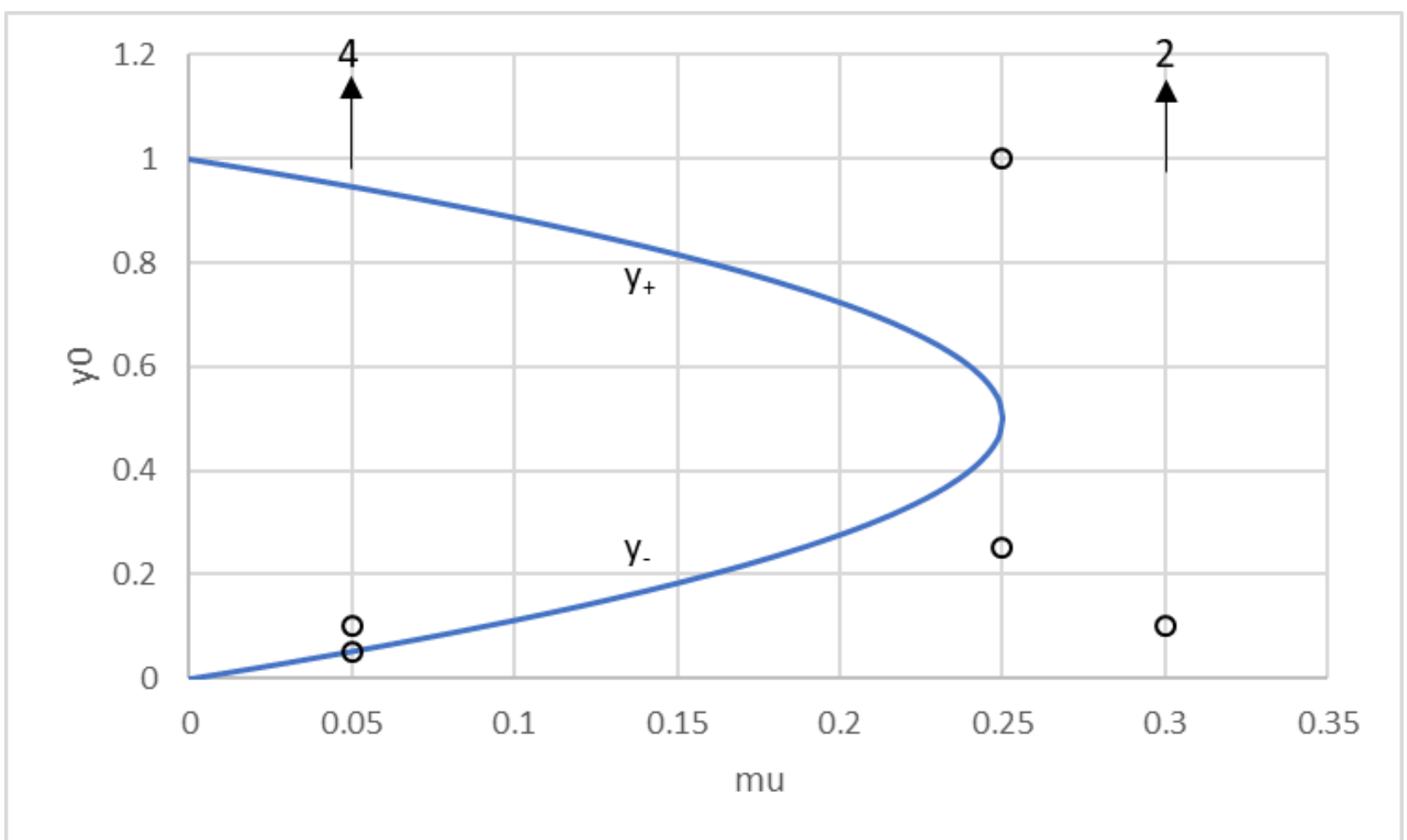

Fig. 1. Phase diagram of $y_0$ vs. $\mu$. Markers indicate the coordinates $(\mu, y_0)$ of the calculation scenarios. Arrows indicate that the points with ordinates $y_0 = 2$ and 4 extend beyond the graph. For $\mu = 0.05$ the lower point is located not on the parabola itself, but slightly below: $y_0 = 0.05 < y_- = 0.053$.

*4.4. Development scenarios*

Let us consider several scenarios for the development of an intelligent system with different values of the parameters $\mu$ and $y_0$ (the scenario positions on the phase diagram is indicated by circles).

**Mode 1** (Fig. 2), $\mu = bc/a < 1/4$. The value $\mu = 0.05$ is taken. Autocatalysis exceeds inhibition: $a > 4bc$. Scenarios with values $y_0$ = 0.05, 0.1 and 4 from three characteristic ranges corresponding to low ($y_0 < y_-$), medium ($y_- < y_0 < y_+$) and high ($y_0 > y_+$) initial rate are calculated, where $y_- = 0.053,\ y_+ = 0.95$.

a) At a low initial rate $y_0 = 0.05 < y_-$, the system cannot overcome inhibition despite the high intensity of autocatalysis. The rate gradually decreases to zero $y \to 0$, meaning knowledge production ceases. The knowledge volume stops at level $x \to (1/k) \ln p$. Such a system is unviable.
b) The opposite situation occurs when the initial rate $y_0 = 0.1$ is in the midrange $y_- < y_0 < y_+$. Knowledge production begins to increase slowly, acceleration reaches a maximum at the inflection point, after which acceleration declines to zero, and the rate curve reaches a plateau at $y \to y_+ = 0.95$, which corresponds to a constant rate of knowledge production.
c) In a high initial rate scenario $y_0 = 4 > y_+$, the system cannot maintain such a high level: the rate decreases to a certain non-zero level (the same as in the previous case: $y \to y_+ = 0.95$), ensuring the growth of knowledge production at a constant rate.

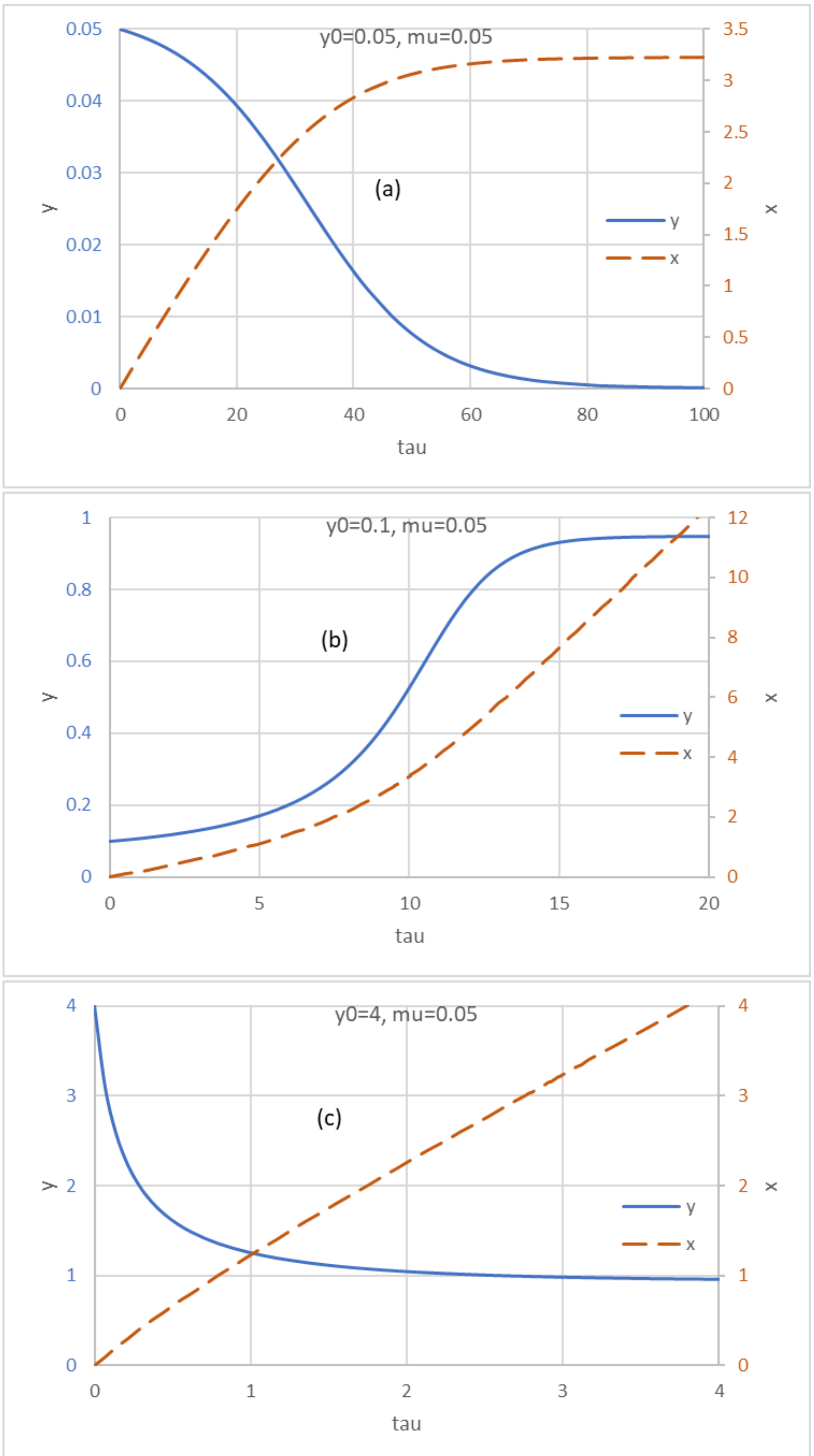

Fig. 2. Knowledge production rate ($y$) and knowledge volume ($x$).
Mode 1: Autocatalysis exceeds inhibition: $a > 4bc,\ \mu < 1/4$.
(a) Low initial rate: $y_0 < y_-$. The rate drops to zero, knowledge ceases to grow, development stops, and the system collapses.
(b) Medium initial rate: $y_- < y_0 < y_+$. The rate increases, reaching a high plateau, knowledge accumulates rapidly, and the system progresses.
(c) High initial rate: $y_0 > y_+$. The rate drops to a low (non-zero) level, the system degrades compared to the initial state, knowledge is produced slowly, and development is maintained at a low rate.

**Mode 2** (Fig. 3), $\mu = 1/4$. Inhibition and autocatalysis are balanced: $a = 4bc$. There are two characteristic ranges with low and high initial rates: $y_0 < 1/2$ and $y_0 > 1/2$.

a) A low value of $y_0 = 0.25 < 1/2$ does not ensure survival, as in the first scenario in Mode 1. The rate of knowledge production decreases to zero ($y \to 0$) with fatal consequences for the system. The knowledge volume stops at level $x \to 4y_0/(1 - 2y_0)$.
b) A high value of $y_0 = 1 > 1/2$ cannot be maintained, however the rate does not fall to zero, but decreases to a final limit: $y \to 1/2$, by analogy with the third scenario in Mode 1. The accumulation of knowledge continues at a constant rate.

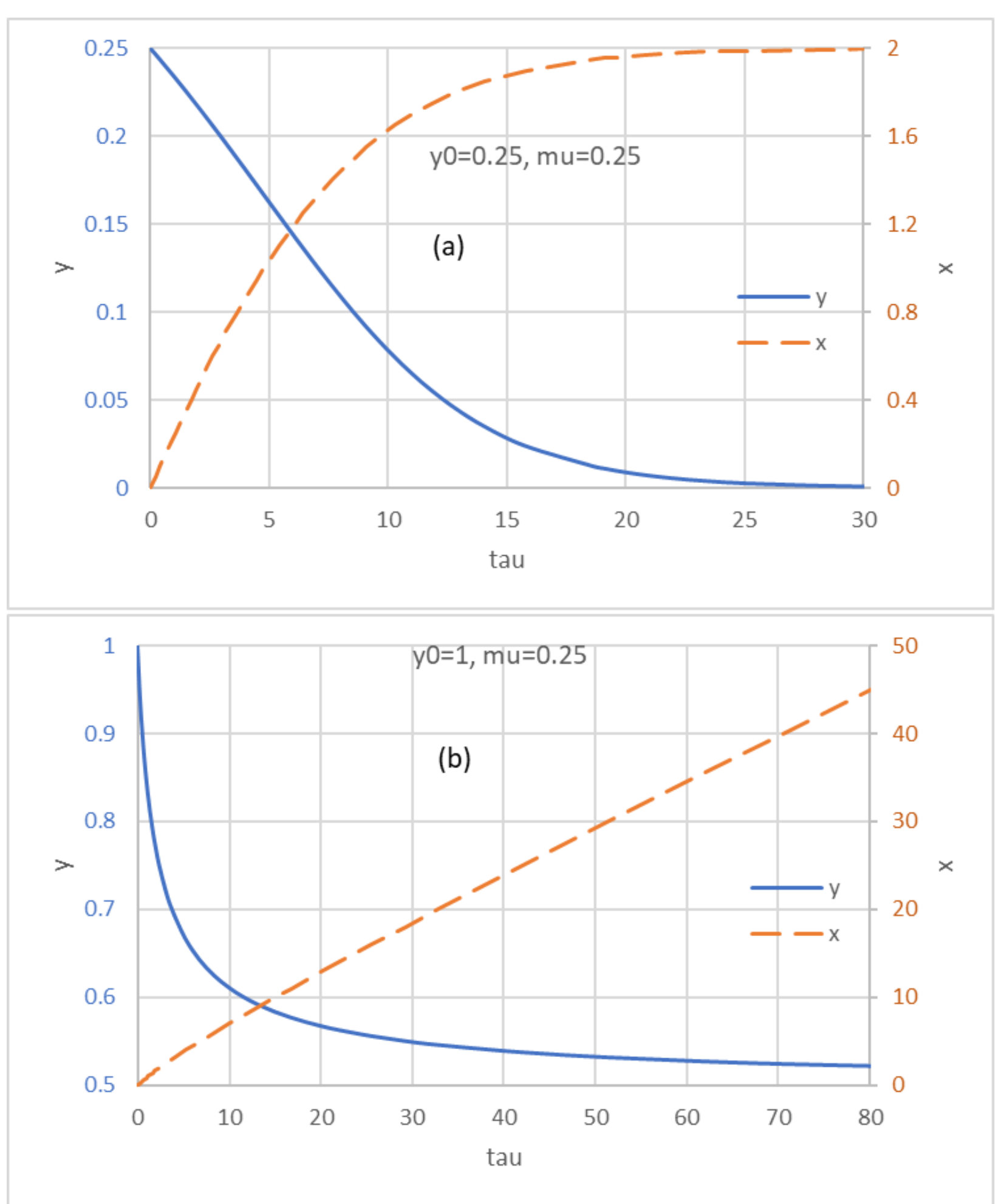

Fig. 3. Mode 2: Balance between autocatalysis and inhibition: $a = 4bc,\ \mu = 1/4$.
(a) Low initial rate: $y_0 < 1/2$. Over time, the rate drops to zero, and the system collapses.
(b) High initial rate: $y_0 > 1/2$. The rate drops to a final level of $y = 1/2$. The system continues to evolve at this rate in the late stage.

**Mode 3** (Fig. 4), $\mu > 1/4$. The value $\mu = 0.3$ is taken. Autocatalysis is weak compared to inhibition: $a < 4bc$. Both scenarios with a small ($y_0 = 0.1$) and a large ($y_0 = 2$) initial rate lead to a drop in rate to zero, and the system collapses.

Thus, *there are only three groups of scenarios with favorable consequences for the intelligent system*. The parameters $\mu$ and $y_0$ for these groups are in the following ranges:

- Mode 1: $\mu < 1/4$, two groups of scenarios: $y_- < y_0 < y_+$ and $y_0 > y_+$.
- Mode 2: $\mu = 1/4$, one group: $y_0 > 1/2$.

(Here we can add trivial scenarios at the boundaries of the specified ranges for $y_0$, in which the rate does not change at all from the beginning to the end of the process: $y(x) = y_0,\ x = y_0\tau$.)

All other groups of scenarios are fatal to the system.

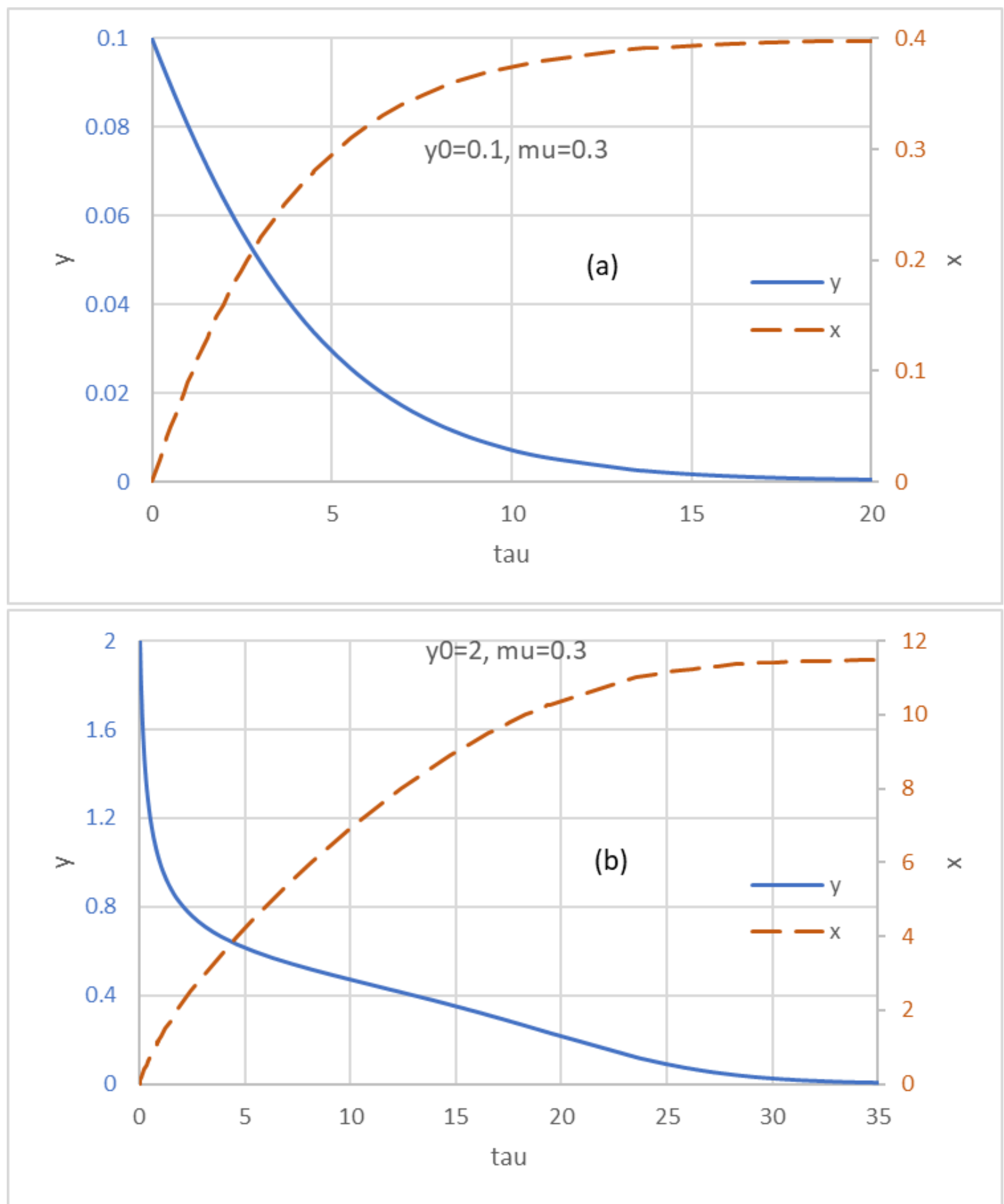

Fig. 4. Mode 3: Inhibition exceeds autocatalysis: $a < 4bc,\ \mu > 1/4$. In both scenarios, the rate decreases to zero, knowledge reaches a finite limit, development stops, and the system collapses.

# 5. Slow subsystem

*5.1. Model coefficients vs. knowledge*

Accumulating knowledge $q$ promotes an increase in the coefficients $a$, $b$, and $c$. The class of functions that describes the dependence of these coefficients on $q$ must include constant, linear, power, and exponential functions as special cases. These requirements are satisfied by *the deformed Tsallis exponential* (Umarov et al., 2008), which we will write in a form more convenient for our purposes:

$$\mathrm{Exp}_\sigma(x) = \left(1 + \frac{x}{\sigma}\right)^\sigma. \tag{23}$$

The limit $\sigma \to 0^+$ gives 1. For $\sigma \to \infty$ we get the usual exponential function $\exp(x)$. In other cases, we have a power function (in particular, linear at $\sigma = 1$).

Let us apply this class of functions to the model coefficients $a$, $b$, and $c$:

- The development of AI (software, architecture, chips, number of parameters, volume of training, etc.) contributes to the growth of autocatalysis:

$$a(q) = a_0\, \mathrm{Exp}_\alpha(k_a q). \tag{24}$$

- Limited resources and environmental problems hinder development, and this tendency increases with increasing $q$:

$$b(q) = b_0\, \mathrm{Exp}_\beta(k_b q). \tag{25}$$

- As $q$ grows under the influence of AI, information noise increases due to the massive injection of unreliable or simply false information. The number of untested models, dead-end generative hypotheses, outdated software packages, and obsolete industrial drawings also grows, creating a tremendous noise intensity and knowledge loss:

$$c(q) = c_0\, \mathrm{Exp}_\gamma(k_c q). \tag{26}$$

*5.2. Adiabatic approximation*

Slow adaptation is possible for small values of parameters $k_a, k_b, k_c$ or exponents $\alpha, \beta, \gamma$.

According to (24)–(26), the parameter $\mu = bc/a$ is now a function of $q$. For sufficiently small $q$, the constant $\mu$ approximation is valid:

$$\mu \approx \mu_0 = \frac{b_0 c_0}{a_0}. \tag{27}$$

With a large volume of knowledge, the asymptotic dependence is

$$\mu \approx \mu_1 (a_0 q)^{\epsilon}, \;\; \epsilon = \beta + \gamma - \alpha, \tag{28}$$

where $\mu_1$ can be expressed through the parameters in (24)–(26). The exponent $\epsilon$ can have any sign, which opens up three scenarios:

- $\epsilon = 0$: at an advanced stage of development, the parameter $\mu$ again becomes constant, but at a different level ($\mu_1$) compared to the initial one ($\mu_0$) [a coincidence of the levels is possible for special values of the parameters in (24)–(26)].
- $\epsilon > 0$: the parameter $\mu$ continues to increase because inhibition increases faster than autocatalysis. The fate of such a civilization is sad: knowledge production eventually drops to zero, and knowledge stagnates at the level achieved at that point. This is the end: without knowledge production, the system collapses.
- $\epsilon < 0$: the parameter $\mu$ decreases because autocatalysis increases faster than inhibition. Knowledge accumulates at an increasing rate. Civilization prospers.

But the prosperity at $\epsilon < 0$ cannot last forever. It will end when $dv/dq = 0$ is reached due to increasing resource-environmental problems and knowledge loss. This is the inflection point of the $q(t)$ curve. At this point, the system transitions from a hyperbolic "takeoff" to a logistic deceleration due to the inhibitory factors. The adiabatic approximation ceases to apply, and the dynamic equation (6) with variable coefficients (24)–(26) must be solved without simplifications.

*5.3. Long-term adaptation*

Let us consider how the slow, long-term trend of the coefficients $a$, $b$, and $c$ affects the system's dynamics as knowledge accumulates. Adopting the adiabatic approximation, we must substitute expressions for $a$, $b$, and $c$ into the formula $\mu = bc/a$. The problem is that expressions (24)–(26) contain many parameters, complicating the analysis. Such detailed expressions are redundant for a qualitative understanding of the slow subsystem's dynamics. Therefore, we use an interpolation formula for $\mu$, combining the initial value (27) and the asymptotic expression (28):

$$\mu = \mu_0 + \mu_1 x^{\epsilon}. \tag{29}$$

The relationship between dimensional and dimensionless quantities is given by formulas (8) and (9) with a slight modification: $a, b, c$ should be replaced by $a_0, b_0, c_0$.

Substituting (29) into the previously obtained equations for the fast subsystem (see Sections 4.2 and 4.3), we can find $x$ and $y$ as functions of time $\tau$.

The calculation results are presented in Fig. 5 on a linear and logarithmic time scale. The following parameter values are used for the calculations:

$$\mu_0 = 0.05, \ \ \mu_1 = 0.05, \ \ \epsilon = 0.1, \ \ y_0 = 0.5. \tag{30}$$

Figure 5a shows the dynamics of the initial stage of the process on a linear time scale. The knowledge production rate $y$, initially equal to 0.5, increases and seems to reach a plateau. However, continuing the calculations to more distant times reveals that this plateau is illusory, and that in fact there is a long-term downward trend, which is visible in Fig. 5b on a logarithmic time scale. The entire process takes approximately $2.1 \cdot 10^6$ units on a dimensionless scale. The process ends fatally for the system: the rate becomes zero and knowledge production ceases. This result is a consequence of the positiveness of the $\epsilon$ exponent, as discussed in Section 5.2.

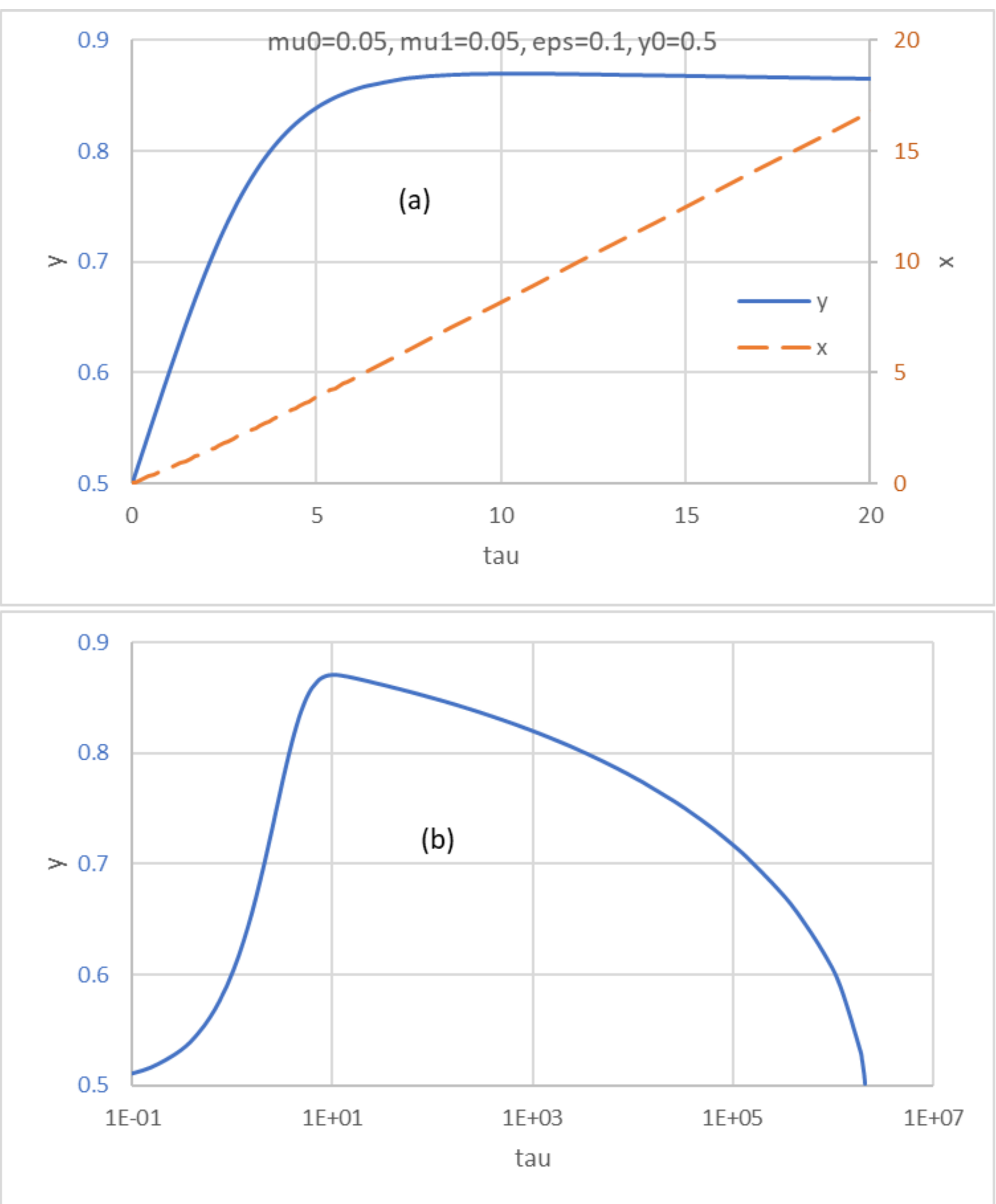

Fig. 5. Dynamics with variable coefficients in the adiabatic approximation $\mu = \mu_0 + \mu_1 x^\epsilon$ for a positive $\epsilon$ exponent. Parameter values are indicated in panel (a).
(a) Initial dynamics: autocatalysis dominates inhibition, the rate of knowledge production increases, reaching a seeming plateau.
(b) Long-term dynamics (logarithmic time scale): inhibition increases, the plateau degenerates, the rate drops to zero, and the system collapses.

## 6. Knowledge-driven population growth

*6.1. Population equation*

According to Malthus (1798), "Population is necessarily limited by the means of subsistence". In other words, the development of new technologies promotes population growth. Since technology is knowledge, a logical connection can be established between knowledge and population size, which was done in the work (Dolgonosov, 2024). The starting point was the knowledge dynamics transformed into population dynamics, in which the dissipation factor (knowledge loss) was expressed through a temperature anomaly caused by $CO_2$ emissions from fossil fuel combustion. Here we will abstract from this effect and consider the direct connection between knowledge dynamics and demography.

AI is currently beginning to be used in knowledge production. At this stage of development, AI is seen as a tool that accelerates this process. Let $w$ be the per capita productivity. It depends on the accumulated knowledge $q$ in accordance with the function (23), as shown in the work (Dolgonosov, 2025). The system as a whole produces knowledge at a rate of

$$v = w(q)N, \tag{31}$$

where $N$ is the population size. After substituting (31) into equation (6) and a series of transformations, we obtain an equation for the function $N(q)$:

$$\frac{dN}{dq} = aN(1 - bN) - \frac{w'}{w}N - \frac{c}{w}, \quad q > q_0, \tag{32}$$

with the initial condition $N(q_0) = N_0$. In (32) all coefficients are, in general, functions of $q$, and the prime denotes the derivative with respect to $q$.

*6.2. Dynamics with constant model coefficients*

Let us consider a basic model with constant coefficients $a$, $b$, $c$, and $w$ and use the results obtained for knowledge dynamics. Equations (31) and (32) can be normalized using the same dimensionless quantities $\tau, x, \mu$ as in (8) and (9), except for two: $y$ and $y_0$, which are now related to the population size:

$$y = bwN, \quad y_0 = bwN_0. \tag{33}$$

Problem (31)–(32) in dimensionless form is as follows:

$$\frac{dy}{dx} = y(1-y) - \mu, \;\; x > 0, \;\; y(0) = y_0 > 0, \tag{34}$$

$$\frac{dx}{d\tau} = y, \;\; \tau > 0, \;\; x(0) = 0. \tag{35}$$

Since $w$ is considered constant ($w' = 0$), the term with $w'$, which is present in (32), drops out of equation (34).

This problem is completely identical to (10) and (17), so all the results obtained in Section 4 apply to demographic dynamics. Only three groups of scenarios with specific intervals of $\mu$ and $y_0$ parameters, described in Section 4.4, are viable, within which the population can either grow and reach a high level or decline and stabilize at a low (non-zero) level. The other groups of scenarios end with the population falling to zero.

The results from Section 5, considering the slow trend of the coefficients as knowledge accumulates (in the adiabatic approximation), are also applicable here. In particular, for a positive $\epsilon$ (see (28)), when autocatalysis is weaker than inhibition ($\alpha < \beta + \gamma$), the system does not survive—the population drops to zero. For zero $\epsilon$ (the balance of autocatalysis and inhibition: $\alpha = \beta + \gamma$), a plateau is reached, corresponding to a constant population. For a negative $\epsilon$ (autocatalysis exceeds inhibition: $\alpha > \beta + \gamma$), the population continues to grow. In fact, growth eventually stops if we solve the dynamics equation with coefficients (24)–(26).

## 7. Discussion. Impact of AI on knowledge dynamics

### *7.1. Dynamics of patent applications*

The application of knowledge dynamics can be illustrated by data series on international patent applications filed since 1978 (WIPO, 2026a, b) and scientific preprints since 1991 (arXiv, 2025) (Fig. 6). Clearly, patents and preprints are far from all the knowledge produced. To apply our dynamic model to these data, we must keep in mind that the model describes the accumulation of knowledge over the entire history, long before the establishment of the corresponding repositories. Therefore, applying the model to these data makes sense only after a certain period of initial accumulation of applications and preprints. Comparison of the model with the data shows that this period lasts approximately two decades. Only after this can we consider that the rate of accumulation of storage units is approximately proportional to the total production of knowledge. For patents, the initial period is 1978–1997. The remaining period (1998–present) is well described by the model, as Fig. 7 shows. The theoretical curve has the following parameters values (definitions are given in (8)):

$$\mu = 0.04, \;\; y_0 = 0.05, \;\; t_0 = 1900, \;\; t_s = 4 \text{ years}.$$

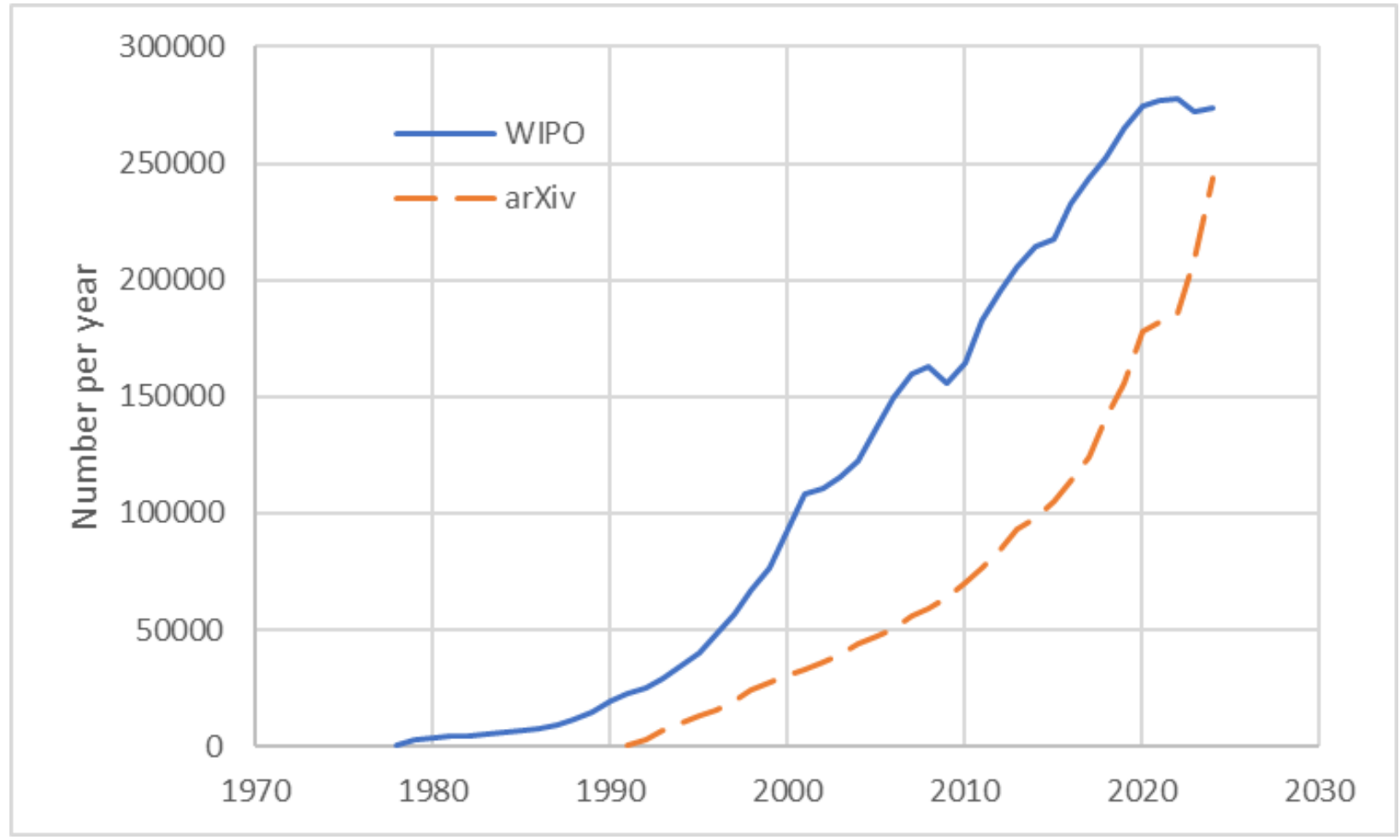

Fig. 6. Annual number of international patent applications received in WIPO and scientific preprints received in arXiv. Data sources: arXiv, 2025; WIPO, 2026a, b.

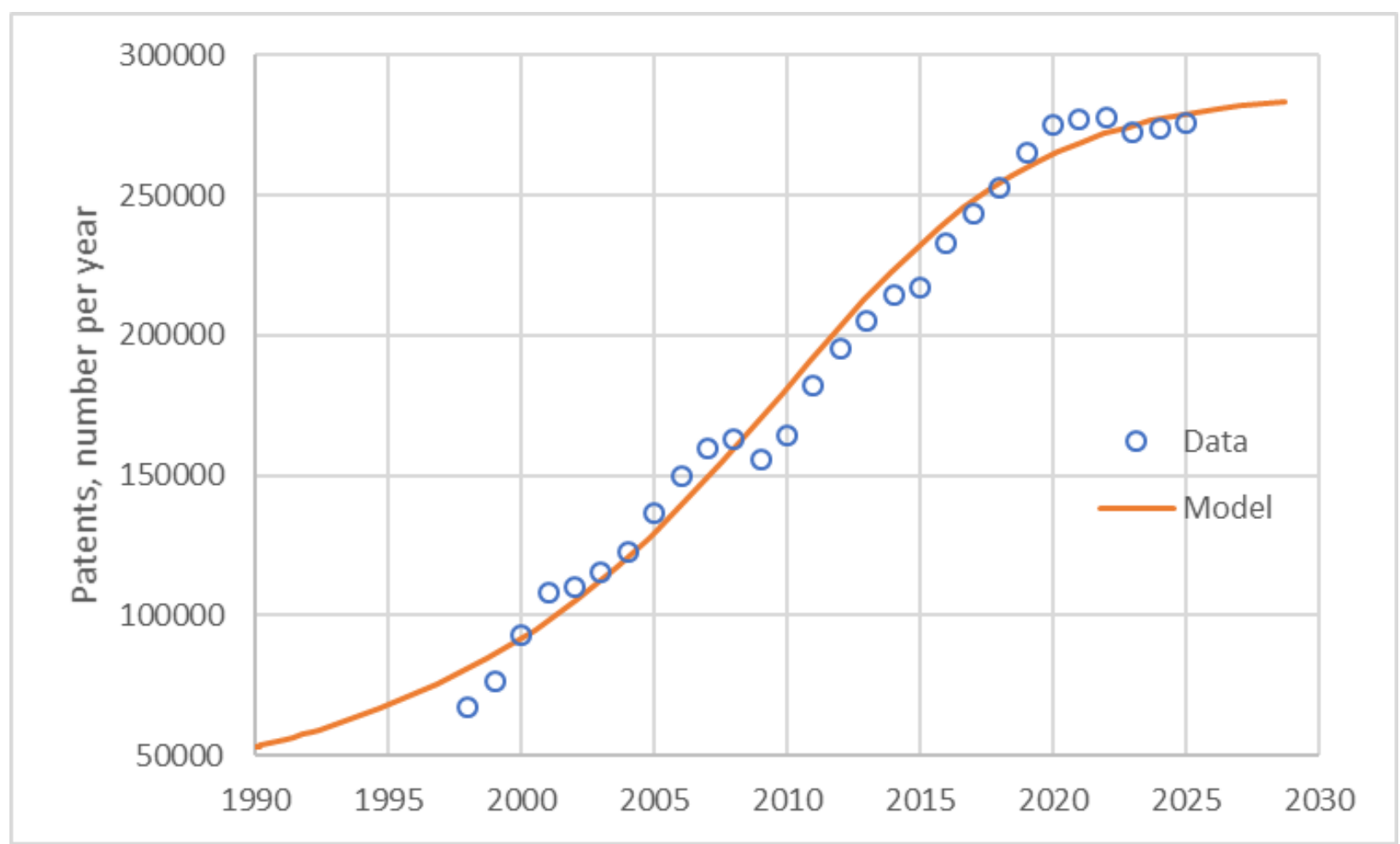

Fig. 7 Annual patent filings at WIPO compared with model calculations. Model parameter values (see definitions in (8) and (9)): $\mu = 0.04,\ y_0 = 0.05,\ t_0 = 1900,\ t_s = 4$ years.

According to (9), the corresponding values of the model coefficients are as follows:

$$a = 0.833 \text{ 1/million},\ b = 3.33 \text{ year/million},\ c = 0.01 \text{ 1/year}.$$

The following physical quantities are associated with the coefficients:

$$t_c = \frac{1}{c},\quad v_b = \frac{1}{b},\quad q_a = \frac{1}{a},\quad v_0 = v_b y_0.$$

Their values for this scenario are:

$$t_c = 100 \text{ years},\ v_b = 0.3 \text{ million/year},\ q_a = 1.2 \text{ million},\ v_0 = 0.015 \text{ million/year},$$

where $t_c$ is the dissipation time scale, $v_b$ is the threshold rate (just about the plateau level on the curve), $q_a$ is the scale of number of patent applications, and $v_0$ is the initial rate (at time $t_0$). For $t_c$, an ordinal estimate of the technology's lifetime is adopted.

Around 2020, the rate of filings reaches a plateau. By this time, AI has not yet reached a level acceptable for practical application, particularly in engineering, and therefore has not become widely adopted. This implies that the observed patent curve formed without AI. As AI develops and is integrated into engineering, labor productivity in this area will increase. Then, the patent curve will resume its upward trend. This behavioral change can be characterized as *a phase transition* from one state (without AI) to another (with AI). To understand how this looks formally, let's turn to the model, specifically formulas (24)–(26) for the coefficients as functions of knowledge volume.

### *7.2. Three eras of knowledge dynamics*

Let us consider how model coefficients change as civilization develops. In terms of knowledge production, three eras can be distinguished: pre-information, information, and the AI era.

*The pre-information era* lasted until approximately the 1960s–1970s, when computers began to find widespread use in various fields. This dating can be judged by the demographic curve of humanity (Dolgonosov, 2016). Before this point, population growth over the last two thousand years is adequately described by a hyperbolic law with a constant autocatalytic coefficient: $a(q) = a_0$. Formula (24) allows for a constant coefficient as a special case when $k_a = 0$ or $\alpha = 0$.

*The information era* is measured from the end of the previous era to the early 2020s (the spread of AI). Using time series of patents, scientific articles, and books, Dolgonosov (2025) shows that productivity grows with knowledge accumulation according to a linear, power, or exponential law—all captured by formula (24). Compared to the previous era, the parameters have other values, indicating a phase transition from one state (without computers) to another (with computers).

*The AI era*: its formation is taking place at the present time. Engineering is still in the early stages of this transition, as evidenced by the stagnation in annual patent filings at WIPO (Fig. 6). As AI tools are implemented, renewed growth is expected in this field. The parameters of autocatalysis should change, enabling increased productivity, which will also be a manifestation of the phase transition.

### *7.3. Dynamics of scientific preprints*

What has been said about patents can be transferred to the scientific preprints accumulated in the repository arXiv (2025). A slight decline in activity is observed in the 2021–2022 period (Fig. 6) due to the COVID-19 pandemic. Then, a rapid increase in activity occurs, reaching an even higher value (~34000 in 2024) than before the decline (~22000 in 2020). The explosive growth observed in 2023–2024 is driven by an avalanche of preprints on computer science and generative AI. AI tools began to generate new scientific hypotheses, inducing autocatalytic acceleration (hyperbolic racing $av^2$).

In the scientific field, the development of AI manifested itself earlier than in patents, since such detailed research as in engineering innovations is not always required. Applying the dynamic model to data from the arXiv repository can now yield little due to the insufficient length of the series: the period of initial accumulation (since 1991) lasts approximately two decades, so that there is too short a portion of the series left for processing.

In 2021–2023, a decline in patent activity was also observed, in sync with preprint activity, due to the pandemic (WIPO, 2026a). The previous decline in patent activity in 2008–2009 coincided with the global financial crisis.

### *7.4. The impact of AI on knowledge production*

The impact of AI on the increased rate of knowledge production can be synchronized with the AI development. Using ChatGPT (OpenAI) as an example, we will examine how the key metrics of this neural network change over time. For illustration purposes, we will limit ourselves to the following three metrics:

- Computing power: the number of floating-point operations spent on training the neural network (FLOP).
- The size of the datasets on which the AI is trained (number of tokens). Modern models ingest virtually all available high-quality text content on the internet.
- The number of parameters as an indicator that characterizes the volume of “knowledge” and connections within the neural network.

Figure 8 shows that AI reaches a level suitable for practical use only by 2020. In 2024, there is a surge in demand in the global market (Fig. 9), driven by the release of the more powerful GPT-4 model. The key stages of generative AI market growth are as follows:

- 2018–2021: Implementation of AI in cloud services and data analytics.
- 2022–2023: Release of public generative AI models (ChatGPT).
- 2024–2025: Rapid growth driven by AI adoption in the corporate sector and investments in infrastructure (NVIDIA chips, data centers);
- 2026: The AI market is expected to reach $900 billion (including hardware).

Market size provides an indirect estimate of AI capabilities. The timeline shows that the second half of the 2020s should see an increase in the rate of knowledge production across various fields. While the data is insufficient for quantitative application of the theoretical framework developed here, the qualitative dynamics are adequately conveyed.

Figures 8 and 9 show that autocatalysis ($av^2$) in AI development continues to operate, albeit with some slowdown ($1 - bv$). From the AI's perspective, this autocatalysis is induced by *external factors*, meaning that the neural network code is written by developers, and tasks are set by users. However, autocatalysis induced by *internal factors* is already beginning to manifest itself—the AI itself is starting to write code for its own versions and, over time, will set its own tasks. This will increase the intensity of autocatalysis (coefficient $a$) and accelerate knowledge production. This provides a clear example of how knowledge creates knowledge. It seems that AI is slowly, imperceptibly, but inevitably displacing humans from the knowledge production circuit. However, humans cannot be completely eliminated from this circuit. The increasing role of AI will most likely lead to a symbiosis between AI and humans in the form of a new evolutionary individuality. "Through recursive feedback, where humans shape AI, and AI increasingly shapes human thought and action, AI may acquire a role not as a separate agent, but as a core architectural element of an emerging collective individual" (Rainey, 2023; Rainey, Hochberg, 2025).

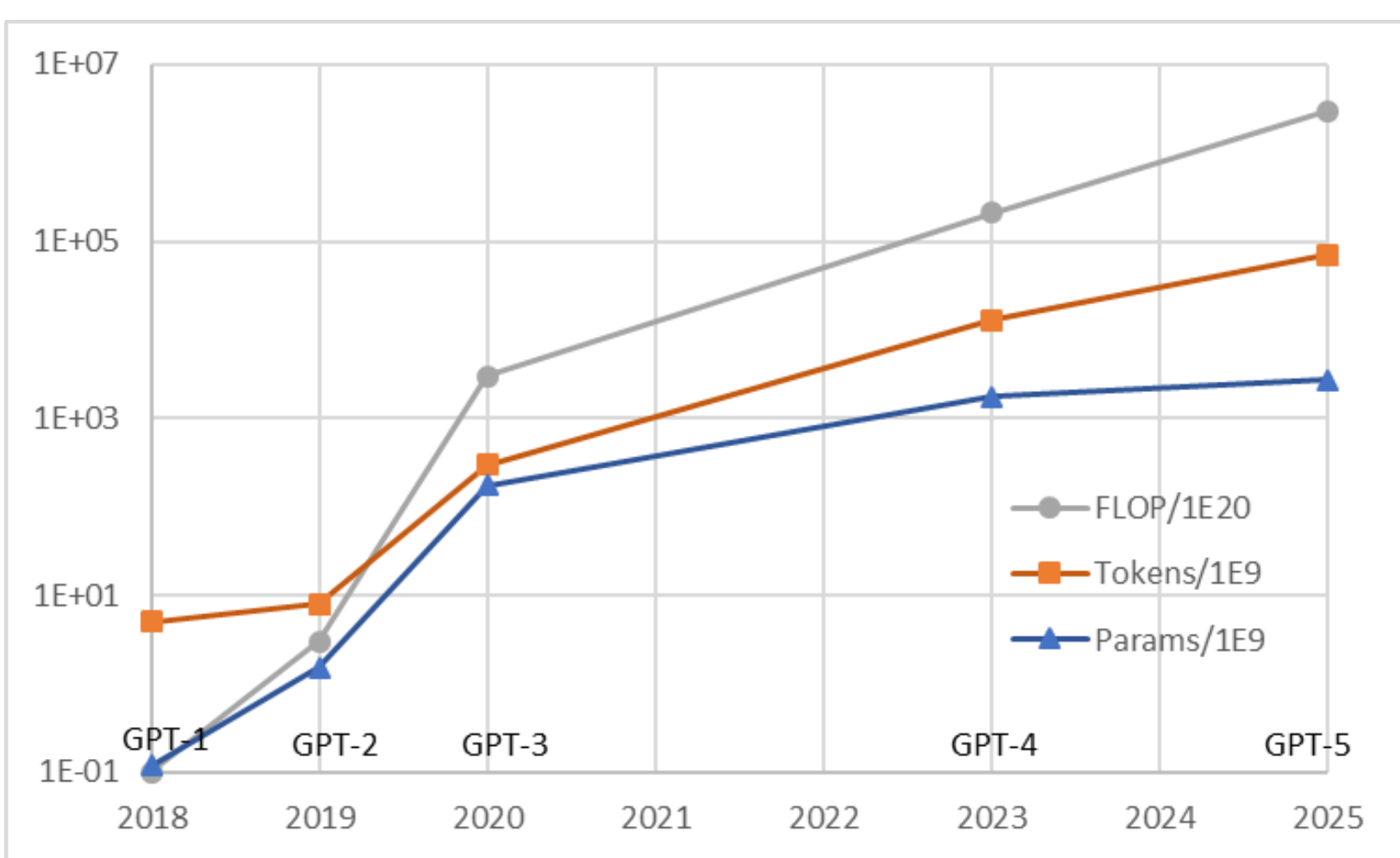

Fig. 8. ChatGPT evolution across three metrics: number of parameters (billions), training volume (tokens, billions), and computing power spent on training (FLOP, $10^{20}$). Data sources: Sevilla et al., 2022; Ho and Berg, 2025; Kumar and Manning, 2025; Rahman et al., 2025.

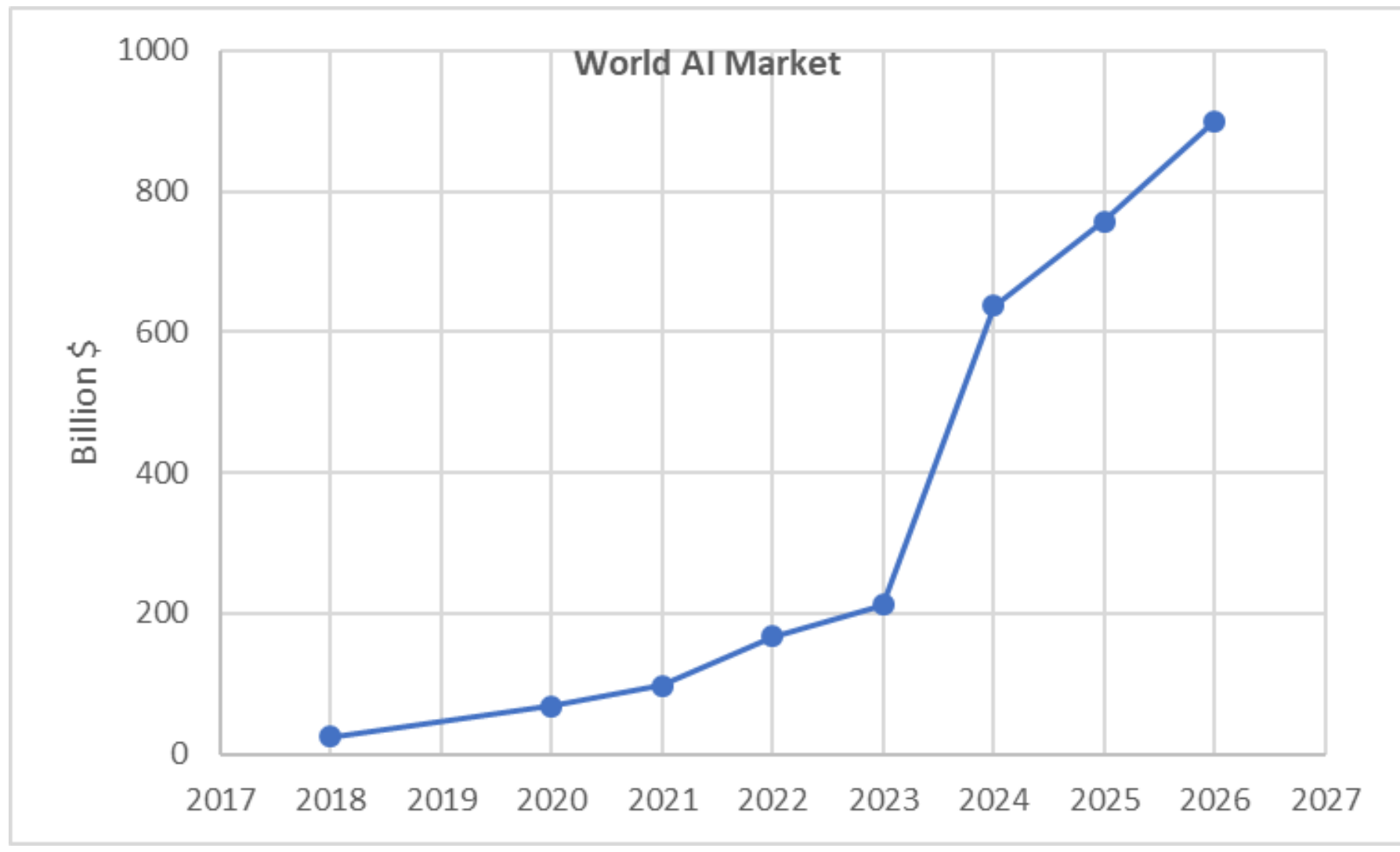

Fig. 9. Global generative AI market by year. Data sources: Statista, 2026; Precedence Research, 2026.

*7.5. The role of resource-environmental brake*

The resource-environmental brake, which corresponds to the factor $(1 - bv) = (v_b - v)/v_b$ in the master equation (2), can be estimated using WIPO patent data (Fig. 6). The graph shows a plateau, which corresponds to an incoming application rate of approximately 0.3 million per year. The plateau is exactly the threshold rate $v_b$ introduced in (3), the reciprocal of which yields the braking coefficient $b = 1/v_b$. This plateau won't last long: as AI is mastered and its power increases, the rate of patent filings will resume growth.

At some point, stagnation will reoccur due to the exhaustion of existing technology capabilities, resulting in a new plateau $v_b$, which will persist until the next technological breakthrough. WIPO's patent activity in microelectronics is already severely limited by the physical lithographic barrier of silicon and the energy consumption of foundries (TSMC, ASML), which slows the rate of actual technology implementation in the real world (Park et al., 2023).

The development of a new technology will dramatically increase the rate of knowledge production, changing the system's state, which represents a phase transition. Thus, *development occurs through a series of phase transitions*.

There is no sign yet of a plateau for preprints in the arXiv repository that would allow us to estimate the threshold rate $v_b$. Preprints do not require peer review, so it is hard to imagine what could slow their filing in the repository, other than perhaps limited storage capacity or preprint filtering, which, however, defeats the purpose of arXiv.

Journal articles are a different matter. For them, the main obstacle can be the increasing delay in peer review due to a large influx of manuscripts. This will lead to stagnation with a corresponding $v_b$ value. Incorporating advanced neural networks into the peer review process can increase the rate of article publication. However, the opposite effect is also possible due to a significant rejection of low-quality material.

In general, *verification* will become the scarcest resource in knowledge production. If a neural network produces a thousand theories in seconds, humanity will spend a hundred years just trying to verify which one won't lead to disaster. Therefore, various AI systems will have to be used to verify theories. *Over time, a critical point will be reached where the function of verification and "comprehension" (integration into the overall world picture) will largely shift from humans to AI.* Knowledge will be detached from its only biological carrier. Humans will cease to be the system's "bottleneck".

*7.6. Barriers to knowledge production*

The growth slowdown caused by various barriers is summarized in the braking coefficient $b$. The rate of knowledge production is directly dependent on AI development. The indicators shown in Figures 8 and 9 demonstrate the growth in power and demand for AI. The growth slowdown here may be due to the presence of three barriers:

1. *Energy barrier.* There is a theoretical minimum energy required to change one bit of information—Landauer's limit (Bennett, 2003). Even if AI becomes as efficient as possible, producing terabytes of knowledge will require a huge amount of energy. When AI begins to consume a significant share of the world's electricity, the rate of knowledge production will slow down.
2. *Thermal barrier.* AI training and operation require massive computations, which generate significant heat. If AI-powered knowledge production reaches a planetary scale, the planet's ability to dissipate this heat will become the limiting factor preventing local or global atmospheric overheating.
3. *The data density barrier*. If AI takes on the function of embedding knowledge into a world picture, it will need to store that picture. Describing it will require a huge number of parameters, reaching the physical limits of data density at the atomic level.

Barriers of the first and third types are currently postponed to the future. The immediate brake leading to stagnation will likely be barrier of the second type—thermal—as climate change caused by greenhouse gas emissions is already being observed. Simply switching to carbon-free energy production is a temporary palliative, as further capacity expansion (especially for AI development) will still lead to thermal heating of the atmosphere. Overcoming the thermal barrier will require a new technological breakthrough aimed at shifting energy-intensive production off-planet.

### *7.7. The role of knowledge loss*

Knowledge loss ($-cv$) is directly related to the concept of *information entropy*: knowledge can become outdated, irrelevant, or simply drown in information noise if the system devotes too few resources to filtering information. Recently, the intensity of noise has increased exponentially due to the widespread dissemination of unreliable or deliberately false information online.

According to recent research (Park et al., 2023), the destructiveness index (CD index) of new patents and publications is declining. This means that despite the enormous number of publications ($q$), the actual breakthrough effect per unit of knowledge production rate ($v$) is declining—the system is expending too much effort on creating irrelevant information or repeating already known facts, which manifests itself as knowledge dissipation.

The knowledge loss coefficient $c$ has a dimension inverse to time. The value $t_c = 1/c$ characterizes the time of knowledge dissipation for the reasons listed above. If we focus on the lifetime of technologies, then the ordinal estimate will be as follows: $t_c \sim 10^2$ years, as adopted in Section 7.1.

## 8. Conclusion

Any intelligent system has survival as its ultimate goal. Knowledge production is a necessary condition and driving force for achieving this goal. Knowledge enables the development of adequate responses to both external impacts and internal challenges, maintaining the system's homeostasis.

1. The article formulates the master equation for knowledge dynamics. The driving force behind knowledge production is autocatalysis, which essentially means that existing knowledge catalyzes the creation of new knowledge. Resource scarcity and information loss inhibit this process.

2. A detailed analysis of the basic version of the dynamic equation was conducted, in which the coefficients of autocatalysis, resource-environmental braking, and information loss are held constant. The basic model demonstrates various modes of knowledge production. Each mode is characterized by a specific range of parameter values. The transition from one mode to another represents a phase transition with a change in the system's dynamics.

3. The area on the phase diagram where autocatalysis prevails over inhibitory factors corresponds to the stable development of the system with an increasing rate of knowledge production.

4. Conditions have been identified under which the system cannot maintain a high rate of knowledge production. The rate drops to a low (non-zero) level. The system degrades but survives because knowledge production continues, albeit at a reduced rate.

5. At a low initial rate, the system cannot survive: the rate drops to zero and knowledge production ceases.

6. A situation where autocatalysis is weaker than inhibition is always fatal for the system, regardless of the initial rate. Over time, the rate drops to zero, leading to system collapse.

7. The case is considered where all factors—autocatalysis, resource scarcity and information dissipation—undergo long-term changes with knowledge growth. Asymptotic analysis allows us to understand under what conditions: 1) the system transitions from one stable state to another, 2) the system continues to develop at an increasing rate, and 3) the rate of knowledge production drops to zero, and the system does not survive.

8. There is a correlation between knowledge and population. This allows us to apply the results obtained to population dynamics, identifying areas in the parametric space where development is stable and where it poses a threat to the survival of civilization.

9. The dynamics of knowledge production are applied to the analysis of patent filing rates at WIPO. Values for the model parameters are determined. A plateau in patent filings and the possible impact of AI on subsequent trends are discussed. Patent application and scientific preprint dynamics are compared with key events in the short history of AI development.

We have developed a mathematical framework for the macrodynamics of knowledge. This is no longer simply an increase in knowledge production, but a balance between a driving force (autocatalysis), a counteracting force (resource-environmental braking), and knowledge loss (dissipation of meanings).

Knowledge is no longer a linear accumulation of facts. It is becoming "intelligent capital" that invests itself. AI shortens the time between a hypothesis being formulated and its verification. Every new discovery made with AI instantly becomes the basis for thousands more, creating a "vertical takeoff" of progress.

**Author information**

Boris M. Dolgonosov, Ph.D. (1979, Lomonosov Moscow State University), D.Sc. (1997, Russian Academy of Sciences), Head of the Ecological modeling laboratory, Institute of Water Problems, Russian Academy of Sciences, Moscow — until 2014. Since 2014: Independent researcher, Haifa, Israel. The author of more than 160 scientific publications including two monographs:

Dolgonosov, B. M., 2009. Nonlinear Dynamics of Ecological and Hydrological Processes. Moscow: Editorial URSS, Librokom. 440 pp. (in Russian).

Dolgonosov, B. M., Gubernatorova, T. N., 2011. Mechanisms and Kinetics of Organic Matter Destruction in Aquatic Environments. Moscow: Editorial URSS, Krasand. 208 pp. (in Russian).

E-mail: borismd31@gmail.com.